\documentclass[12pt]{iopart}

\usepackage{epsf}

\def\agt{\mathrel{\raise.3ex\hbox{$>$}\mkern-14mu\lower0.6ex\hbox{$\sim$}}}
\def\alt{\mathrel{\raise.3ex\hbox{$<$}\mkern-14mu\lower0.6ex\hbox{$\sim$}}}

\newcommand{\beq}{\begin{equation}}
\newcommand{\eeq}{\end{equation}}
\newcommand{\beqn}{\begin{eqnarray}}
\newcommand{\eeqn}{\end{eqnarray}}
\newcommand{\pa}{\partial}

\newcommand{\varep}{\varepsilon}

\newcommand{\brr}{{\mbox{\boldmath$r$}}}

\begin{document}

\title{Merger of black hole-neutron star binaries in full general relativity}

\author{Masaru Shibata$^1$ and Koji Ury\=u$^2$}

\address{$^1$Graduate School of Arts and Sciences, 
University of Tokyo, Komaba, Meguro, Tokyo 153-8902, Japan\\
$^2$ Department of Physics, University of Wisconsin-Milwaukee, 
P.O. Box413, Milwaukee, WI 53201, USA}

\begin{abstract}
We present our latest results for simulation for merger of black hole
(BH)-neutron star (NS) binaries in full general relativity which is 
performed preparing a quasicircular state as initial condition. The BH
is modeled by a moving puncture with no spin and the NS by the
$\Gamma$-law equation of state with $\Gamma=2$ and corotating velocity
field as a first step.  The mass of the BH is chosen to be $\approx
3.2 M_{\odot}$ or $4.0M_{\odot}$, and the rest-mass of the NS $\approx
1.4 M_{\odot}$ with relatively large radius of the NS $\approx 13$--14
km. The NS is tidally disrupted near the innermost stable orbit but
$\sim 80$--90\% of the material is swallowed into the BH and resulting
disk mass is not very large as $\sim 0.3M_{\odot}$ even for small BH
mass $\sim 3.2M_{\odot}$.  The result indicates that the system of a
BH and a massive disk of $\sim M_{\odot}$ is not formed from
nonspinning BH-NS binaries irrespective of BH mass, although a disk of
mass $\sim 0.1M_{\odot}$ is a possible outcome for this relatively
small BH mass range as $\sim 3$--4$M_{\odot}$. Our results indicate
that the merger of low-mass BH and NS may form a central engine of
short-gamma-ray bursts.
\end{abstract}

\section{Introduction}

Merger of black hole (BH)-neutron star (NS) binaries is one of likely
sources of kilo-meter size laserinterferometric gravitational wave
detectors.  Although such system has not been observed yet in contrast
to NS-NS binaries, statistical studies based on the stellar evolution
synthesis suggest that the merger will happen more than 10\% as
frequently as the merger of binary NSs \cite{grb1,BHNS}. Thus, the
detection of such system will be achieved by laserinterferometers in
near future. This motivates theoretical studies for the merger of
BH-NS binaries. 

According to a study based on the tidal approximation (which is
referred to as a study for configuration of a Newtonian star in
circular orbits around a BH in its relativistic tidal field; e.g.,
\cite{F,shibata,WL,ISM,FBSTR}), the fate is classified into two cases,
depending on the mass ratio $q \equiv M_{\rm NS}/M_{\rm BH}$, where
$M_{\rm BH}$ and $M_{\rm NS}$ denote the masses of BH and NS,
respectively. For $q \alt q_c$, the NS of radius $R$ will be swallowed
into the BH horizon without tidal disruption before the orbit reaches
the innermost stable circular orbit (ISCO) \cite{WL,ISM}. On the other
hand, for $q \agt q_c$, NS may be tidally disrupted before plunging
into BH. Here, the critical value of $q_c$ depends on the BH spin and
equation of state (EOS) of NS. For the nonspinning case with stiff
EOSs, $q_c \approx 0.3$--$0.35(R/5M_{\rm NS})^{-3/2}$ and for the case
that spin of a BH aligns with the orbital angular momentum, $q_c$ can
be larger \cite{ISM}. (Throughout this paper, we adopt the geometrical
units $c=G=1$.) 

The tidal disruption has been studied with great interest because of
the following reasons. (i) Gravitational waves at tidal disruption
will bring information about the NS radius since the tidal disruption
limit depends sensitively on it \cite{valli}. The relation between the
mass and the radius of NSs may be used for determining the EOS of high
density matter \cite{lindblom}. (ii) Tidally disrupted NSs may form a
massive disk of mass $\sim 0.1$--$1M_{\odot}$ around BH if the tidal
disruption occurs outside the ISCO. Systems consisting of a rotating
BH surrounded by a massive, hot disk have been proposed as one of
likely sources for the central engine of gamma-ray bursts (GRBs) with
a short duration \cite{grb2}, and hence, merger of low-mass BH and NS
is a candidate.

However, the scenario based on the tidal approximation studies may be
incorrect since gravitational radiation reaction and
self-gravitational effects of NS to the orbital motion are
ignored. Radiation reaction shortens the time available for tidally
disrupting NSs. The gravity of NS could increase the orbital radius of
the ISCO and hence the critical value of the tidal disruption, $q_c$,
may be larger in reality. Miller \cite{CMiller} estimates these
ignored effects and suggests that NSs of canonical mass and radius
will be swallowed into BH without tidal disruption. Moreover, NSs are
described by the Newtonian gravity in the tidal approximation. If we
treat it in general relativity, the self-gravity is stronger and hence
tidal disruption is less likely. 

Tidal disruption of NSs by a BH has been investigated in the Newtonian
\cite{Newton} and approximately general relativistic (GR) simulation
\cite{FBST,FBSTR}. However, the criterion of the tidal disruption will
depend on GR effects as mentioned above, and hence, a simulation in
full general relativity is obviously required (see \cite{loffler} for
an effort). In \cite{SU06}, we present our first numerical results for
fully GR simulation, performed by our new code which has been improved
from previous one \cite{STU0,STU}; we handle an orbiting BH 
adopting the moving puncture method, which has been recently developed
by two groups \cite{BB2} (see also \cite{BB4} for detailed calibration
of this method). As the initial condition, we prepare a
quasicircular state computed in a new formalism described in
Sec. 2. Focusing particularly on whether NSs of realistic mass and
radius is tidally disrupted to form a massive disk around nonspinning
BHs, we illustrate that a disk with mass $\sim M_{\odot}$ is an
unlikely outcome for plausible values of NS mass and radius and for BH
mass greater than $3M_{\odot}$, although a disk of mass of
$O(0.1M_{\odot})$ is possible. 

In this paper, we extend previous study; we perform a simulation for
different BH mass from that in \cite{SU06} to find the dependence of
the disk mass on the BH mass.  In addition, gravitational waves are
derived. The paper is organized as follows. In Sec. 2, we describe a
formulation for computing quasicircular states in the moving-puncture
framework. Section 3 presents some of numerical results for the
quasicircular orbits. In Sec. 4, we report numerical results of 
simulations for the merger of BH-NS binaries. Section 5 is devoted to a
summary and discussion.

\section{Formalism for a quasicircular state} 

Three groups have worked in computing quasicircular states of BH-NS
binaries \cite{Miller,TBFS,GRAN}. However, this field is still in an
early stage in contrast to computation for NS-NS binaries (e.g.,
\cite{ULFGS}).  In \cite{SU06}, we proposed a new method for computing
accurate quasicircular states that can be used for numerical
simulation in the moving puncture framework \cite{BB,BB2}. Here, we
briefly describe the formulation. 

Even just before the merger, it is acceptable to assume that BH-NS
binaries are in a quasicircular orbit since the time scale of
gravitational radiation reaction is a few times longer than the
orbital period. Thus, we assume the presence of a helical Killing
vector around the mass center of the system, $\ell^{\mu}=(\pa_t)^{\mu}
+\Omega (\pa_{\varphi})^{\mu}$, 
where the orbital angular velocity $\Omega$ is constant.  

In the present work, we assume that the NS is corotating around the
mass center of the system for simplicity.  Irrotational velocity field
is believed to be more realistic for BH-NS binaries \cite{KBC}. The
work for the irrotational case will be reported in the future paper
\cite{US07}.  The assumption of corotating velocity field in the
helical symmetric spacetime yields the first integral of the Euler
equation, $h^{-1} u^t={\rm const}$,
where $h$ is specific enthalpy defined by $1+\varepsilon+P/\rho$,
and $\varep$, $P$, and $\rho$ are specific internal energy, pressure,
and rest-mass density, respectively. In the present work, we adopt the
$\Gamma$-law EOS with $\Gamma=2$; $P=\rho\varep=\kappa \rho^2$
with $\kappa$ an adiabatic constant. $u^{\mu}$ denotes the four
velocity and $u^t$ its time component. Assumption of corotation
implies $u^{\mu}=u^t \ell^{\mu}$ and thus $u^r=u^{\theta}=0$. 

For a solution of geometric variables of quasicircular orbits, we
adopt the conformal flatness formalism for three-geometry. In this
formalism, the solution is obtained by solving Hamiltonian and
momentum constraint equations, and an equation for the time slicing
condition which is derived from $K_k^{~k}=0$ where $K_{ij}$ is the
extrinsic curvature and $K_k^{~k}$ its trace \cite{IWM}. 
Using the conformal factor $\psi$, the rescaled tracefree
extrinsic curvature $\hat A_{i}^{~j} \equiv \psi^6 K_{i}^{~j}$, and
weighted lapse $\Phi\equiv \alpha\psi$ where $\alpha$ is the lapse
function, these equations are respectively written 
\beqn
&& \Delta \psi = -2\pi \rho_{\rm H} \psi^5 -{1 \over 8}
\hat A_{i}^{~j} \hat A_{j}^{~i}\psi^{-7}, \label{ham2} \\
&& \hat A^{~j}_{i~,j} = 8\pi J_i \psi^6,\label{mom2}
\\
&& \Delta \Phi = 2\pi \Phi \Big[\psi^4 (\rho_{\rm H} + 2 S)
+{7 \over 16\pi} \psi^{-8}\hat A_{i}^{~j} \hat A_{j}^{~i}\Big],
\label{alpsi}
\eeqn
where $\Delta$ denotes the flat Laplacian, 
$\rho_{\rm H}=\rho h (\alpha u^t)^2-P$, $J_i=\rho h u_i$, 
and $S=\rho h [(\alpha u^t)^2-1]+3P$. 

We solve these equations in the framework of the puncture BH
\cite{BB,BB2,hannam}. Assuming that the puncture is located at $\brr_{\rm P}$, 
we set $\psi$ and $\Phi$
\beqn
\psi=1+{M_{\rm P} \over 2 r_{\rm BH}} + \phi
\ \ \mbox{and} \ \  
\Phi=1 - {C \over r_{\rm BH}} + \eta,
\eeqn
where $M_{\rm P}$ and $C$ are positive constants, 
and $r_{\rm BH}=|x^k_{\rm BH}|$ ($x^k_{\rm BH}=x^k-x^k_{\rm P}$).
Then elliptic equations for functions $\phi$ and $\eta$ are derived.
The constant $M_{\rm P}$ is arbitrarily given, while 
$C$ is determined from the virial relation (e.g., \cite{vir})
\beqn
\oint_{r \rightarrow \infty} \pa_i \Phi dS^i=
-\oint_{r \rightarrow \infty} \pa_i \psi dS^i=2\pi M, 
\eeqn
where $M$ is the ADM mass. The mass center is determined from the
condition that the dipole part of $\psi$ at spatial infinity is
zero.
In this method, the region with $\alpha < 0$ exists.  However,
this does not cause any pathology in the initial value problem. 

Equation (\ref{mom2}) is rewritten setting 
\beqn
\hat A_{ij}(=\hat A_i^{~k}\delta_{jk})
=W_{i,j}+W_{j,i}-{2 \over 3}\delta_{ij} \delta^{kl}
W_{k,l}+K^{\rm P}_{ij},\label{hataij}
\eeqn
where $K^{\rm P}_{ij}$ denotes the weighted
extrinsic curvature associated with linear momentum of a puncture BH; 
\beqn
K^{\rm P}_{ij}={3 \over 2 r_{\rm BH}^2}\biggl(n_i P_j +n_j P_i
+(n_i n_j-\delta_{ij}) P_k n_k \biggr). 
\eeqn
Here, $n^k=n_k=x^k_{\rm BH}/r_{\rm BH}$. $P_i$ denotes linear momentum of
the BH, determined from the 
condition that the total linear momentum of system should be zero; 
\beqn
P_i=-\int J_i \psi^6 d^3x. \label{P}
\eeqn
The RHS of Eq. (\ref{P}) denotes the total linear momentum
of the companion NS.
Then, the total angular momentum of the system is derived from
\beqn
J=\int J_{\varphi} \psi^6 d^3x + \epsilon_{zjk} r_{\rm P}^j \delta^{kl}P_l. 
\eeqn

The elliptic equation for $W_i(=W^i)$ is 
\beqn
\Delta W_i + {1 \over 3}\pa_i \pa_k W^k=8\pi J_i \psi^6. \label{weq}
\eeqn
Denoting $W_i=7 B_i - (\chi_{,i}+B_{k,i} x^k)$ where $\chi$ and $B_i$
are auxiliary functions \cite{gw3p2}, Eq. (\ref{weq}) is decomposed as
two linear elliptic equations 
\beq
\Delta B_i = \pi J_i \psi^6~~{\rm and}~~\Delta \chi= -\pi J_i x^i \psi^6.
\eeq

Computing BH-NS binaries in a quasicircular orbit requires to
determine the shift vector even in the puncture framework. This is
because $u_i$ has to be obtained [it is derived from $u_k=\delta_{ki}
u^t \psi^4 (v^i + \beta^i)$ where $v^i=\Omega \varphi^i$].  
The relation between 
$\beta^i$ and $\hat A_{ij}$ is written 
\beqn
\delta_{jk} \pa_i \beta^k +\delta_{ik} \pa_j \beta^k
-{2\over 3}\delta_{ij} \pa_k \beta^k ={2\alpha \over \psi^6} \hat A_{ij}.
\label{eq13}
\eeqn
Operating $\delta^{jl}\pa_l$, an elliptic equation is derived 
\beqn
\Delta \beta^i + {1 \over 3} \delta^{ik} \pa_k\pa_j\beta^j
=2 \pa_j (\alpha \psi^{-6}) \hat A^{ij}+16\pi\alpha J_j \delta^{ij}.
\label{betaeq}
\eeqn
Here for $\hat A_{ij}$, we substitute the relation of
Eq. (\ref{hataij}) (not Eq. (\ref{eq13})). As a result, no singular
term appears in the RHS of Eq. (\ref{betaeq}) and Eq. (\ref{betaeq})
is solved in the same manner as that for $W_i$.

We have computed several models of quasicircular states and found
that the relation between $\Omega$ and $J$
approximately agrees with the 3rd Post-Newtonian relation \cite{Luc}.
This makes us confirm that this approach is a fair way for preparing 
quasicircular states.  We also found that in this method, the shift 
vector at $\brr=\brr_{\rm P}$ automatically satisfies the condition
$\beta^{\varphi}=-\Omega$ within the error of a few \%. This implies
that the puncture is approximately guaranteed to be in a corotating
orbit in the solution. 

\section{Numerical results for quasicircular states}

\begin{table}[tb]
\begin{center}
\caption{Parameters of quasicircular states.  Mass parameter of
puncture, mass of BH, rest-mass of NS, mass, radius, and normalized mass
of NS in isolation, total mass of the system, nondimensional angular momentum
parameter, orbital period in units of $M$, and compactness of the
system defined by $C_o=(M\Omega)^{2/3}$.  Mass of BH is computed from
the area of the apparent horizon $A$ as $(A/16\pi)^{1/2}$. Mass is
shown in units of $M_{\odot}$.  }
\begin{tabular}{ccccccccccc} \hline
 & $M_{\rm P}$ & $M_{\rm BH}$ & $M_{*}$ & $M_{0\rm NS}$ 
& $R$ (km) & $M_{*}/\kappa^{1/2}$ & $M$ & $J/M^2$ & $P_0/M$ & $C_o$ \\ \hline
A & 3.13 & 3.21 & 1.40 & 1.30 & 13.8 & 0.147 & 4.47 & 0.729 & 119 & 0.141 \\ 
B & 3.13 & 3.21 & 1.40 & 1.30 & 13.8 & 0.147 & 4.47 & 0.720 & 110 & 0.150 \\ 
C & 3.93 & 4.01 & 1.40 & 1.30 & 13.0 & 0.151 & 5.26 & 0.645 & 115 & 0.144 \\ \hline 
\end{tabular}
\end{center}
\end{table}

BH-NS binaries in quasicircular orbits have been computed for a wide
variety of models with $q=M_{*}/M_{\rm BH} \approx 0.3$--0.5 where
$M_{*}$ denotes baryon rest-mass of the NS. In the present work, the
compactness of spherical NSs with rest-mass $M_{*}$ is chosen to be
$\approx 0.14$--0.15. In the $\Gamma$-law EOSs, the mass and radius of
NS are rescaled by changing the value of $\kappa$: In the following we
fix the unit by setting that $M_{*}=1.4M_{\odot}$.

In computation, we focus only on the orbit of slightly outside of
ISCO. In Table I, we show the quantities for selected quasicircular
states with $q \approx 0.4$ and 0.33.  For models A and B shown in
Table I, the radius of NS in isolation is $R \approx 13.8$ km, the
gravitational mass in isolation is $M_{0\rm NS} \approx 1.30
M_{\odot}$, and $M_{*}/\kappa^{1/2}=0.147$. For model C, $R \approx
13.0$ km, $M_{0\rm NS} \approx 1.3M_{\odot}$, and
$M_{*}/\kappa^{1/2}=0.151$.  According to theories for NSs based on
realistic nuclear EOSs \cite{EOS}, the radius of NS of $M_{\rm NS}
\approx 1.4 M_{\odot}$ is 11--13 km. Thus the radius chosen here is
slightly larger than that of realistic NSs and is more subject to
tidal disruption.  Model A is that used for the following numerical
simulation and model B is very close to the tidal disruption limit of
approximately the same mass as that of model A, showing that the model
A has an orbit of slightly outside of the tidal disruption limit. This is
also the case for model C. As we show in the next section, tidal
disruption sets in after a small decrease of orbital separation for
models A and C.

The tidal approximation studies suggest that for $q \agt q_*$, NSs
could be tidally disrupted by a BH \cite{ISM}.  Here, in the tidal
approximation, the critical value $q_*$ for $\Gamma=2$ and for
nonspinning BHs is approximately given by
\beq
q_* \equiv 
0.35\biggl({R \over 5M_{\rm NS}}\biggr)^{-3/2}
{(M_{\rm BH}\Omega)^{-1} \over 6^{3/2}}, 
\eeq
and $\Omega=M_{\rm BH}^{-1}/6^{3/2}$ is the angular velocity of the
ISCO around nonspinning BHs.  For models A and B, $q_* \approx 0.32$, and
hence, $q > q_*$.  According to the tidal approximation studies
\cite{WL,ISM}, such NS should be unstable against tidal disruption.
Nevertheless, such equilibrium exists, proving that the tidal
disruption limit in the framework of the tidal approximation does not
give correct answer. Our studies indicate that the critical value
$q_*$ is $\approx 0.43(R/5M_{\rm NS})^{-3/2}[(M_{\rm BH}\Omega)^{-1}
/6^{3/2}]$; tidal disruption of NS is much less likely than in the
prediction by the tidal approximation \cite{WL,ISM}. For the typical
NS of radius $R \sim 5M_{\rm NS}$ and mass $M_{\rm NS} \sim
1.4M_{\odot}$, $M_{\rm BH} \alt 3.3M_{\odot}$ will be necessary for
$(M_{\rm BH}\Omega)^{-2/3}\geq 6$; this implies that canonical NSs
will not be tidally disrupted outside ISCO by most of nonspinning BHs
of mass larger than $\sim 3M_{\odot}$. Tidal disruption occurs only
for NSs of relatively large radius and only for orbits very close to
ISCO.

In this work, the criterion for the tidal disruption is investigated
only for $\Gamma=2$ EOS and for compactness 0.14--0.15. The criterion
is likely to depend on the stiffness of the EOS \cite{ISM} since the
structure of NS depends on it. The criterion should also depend
strongly on the compactness of NS in general relativity which has a
nonlinear nature. In the future paper, we plan to determine the
criterion for a wide variety of EOSs and compactness of NS
\cite{US07}.

\section{Simulation for merger}

Even if tidal disruption of an NS occurs near ISCO, a massive disk may
be formed around the companion BH. To investigate the outcome after merger
and resulting gravitational waveforms, we perform numerical simulation
adopting models A and C.

For the simulation, we initially reset the lapse (i.e., $\Phi$) since
the relation $\alpha \geq 0$ should hold everywhere.
In the present work, $\Phi$ at $t=0$ is given by 
\beqn
\Phi &= & \eta +{1 + 0.1 X^4 \over 1+\sum_{m=1}^{3} X^m +1.1X^4}~,  
\eeqn
where $X=C/r_{\rm BH}$. 
Then, $\alpha=0$ only at puncture and otherwise $\alpha >0$.
Furthermore, for $r_{\rm BH} >C$, the values of $\Phi$ quickly
approach to those of the quasicircular states. 

The numerical code for hydrodynamics is the same as that for
performing merger of NS-NS binaries (a high-resolution central scheme)
\cite{SF,STU}. On the other hand, we change equations for $\alpha$,
$\beta^i$, and $\psi$, and numerical scheme of handling the transport
terms of evolution equations for geometries. For $\alpha$ and
$\beta^i$ we solve \beqn &&(\pa_t -\beta^i \pa_i)\ln \alpha = -2
K_k^{~k},
\label{lapse} \\
&&\pa_t \beta^i=0.75 \tilde \gamma^{ij} (F_j +\Delta t \pa_t F_j),
\label{shift} 
\eeqn
where $\tilde \gamma_{ij}$ is the conformal three-metric and
$F_i=\delta^{jk}\pa_j \tilde \gamma_{ik}$. $\Delta t$ denotes the
time step for the simulation and the second term in the RHS
of Eq. (\ref{shift}) is introduced for stabilization. 
The equation for the conformal factor is also changed to
\beqn
\pa_t \psi^{-6} - \pa_i(\psi^{-6}\beta^i)= (\alpha K_k^{~k}
-2\pa_i\beta^i)\psi^{-6}, 
\eeqn
since $\psi$ diverges at the puncture \cite{BB2}. 

In addition, we have improved numerical scheme for the transport term
of geometric variables $(\pa_t - \beta^i\pa_i )Q$ where $Q$ is one of
the geometric variables: First, we rewrite this term to $\pa_t Q-\pa_i
(Q\beta^i)+Q \pa_i\beta^i$ and then apply the same scheme as in
computing the transport term of the hydrodynamic equations to the
second term (3rd-order piece-wise parabolic interpolation scheme
\cite{STU0}). We have found that for evolving BHs, such
high-resolution scheme for the transport term in the geometric
variables are crucial. This is probably because of the fact that near
punctures, some of geometric variables steeply vary and so is the term
$\beta^i\pa_i Q$. For other terms in the Einstein's equation, we use
the 2nd-order finite differencing as in \cite{STU0,STU}.  (Note that
in the case of nonuniform grid, 4-point finite differencing is adopted
for $Q_{,ii}$ since 3-point one is 1st-order.) After we performed
most of runs, we iterated some of computations with 
a 3rd-order scheme (5-point finite differencing for $Q_{,ii}$)
which is used in \cite{BB2,BB4}. We find that with such scheme, 
convergent results are obtained with a relatively large grid spacing.
However, the results are quanlitatively unchanged and 
the extrapolated results (which are obtained in the limit 
of zero grid spacing; see below) are approximately identical. 

In the simulation, the cell-centered Cartesian, $(x, y, z)$, grid is
adopted to avoid the situation that the location of punctures (which
always stay in the $z=0$ plane) coincides with the grid location. The
equatorial plane symmetry is assumed and the grid size is $(2N, 2N,
N)$ for $x$-$y$-$z$ where $N$ is a constant. Following \cite{SN}, we
adopt a nonuniform grid; in the present approach, a domain of $(2N_0,
2N_0, N_0)$ grid zone is covered with a uniform grid of the spacing
$\Delta x$ and outside the domain, the grid spacing is increased
according to $\xi\tanh[(i-N_0)/\Delta i]\Delta x$ where $i$ denotes
the $i$-th grid point in each direction. $N_0$, $\Delta i$, and $\xi$
are constants.  For model A, $\Delta x/M_{\rm P}$ is chosen to be
$1/8$, $9/80$, $1/10$, 7/80, and 3/40, and for model C, it is $7/120$,
8/120, 9/120, and 1/12. As shown in \cite{BB2}, such grid spacing can
resolve moving punctures.  For model A, $(N, N_0, \Delta i, \xi,
\Delta x/M_p, L/\lambda)$ is chosen to be (160,105,30,4.5,1/8,0.46),
(200,105,30,4.5,1/8,0.78), (200,105,30,5,1/8, 0.83),
(220,125,30,5,9/80, 0.78), (220,125,30,6,1/10, 0.78),
(220,140,30,7,7/80, 0.65), and (220,150,9,3/40, 0.59). Here, $L$ and
$\lambda$ denote the location of the outer boundaries along each axis
and the wavelength of gravitational waves at $t=0$. For model A with
$\Delta x=M_{\rm P}/8$ and $N=160$, we chose other values of $N_0$ and
$\Delta i$, and found that results depend weakly on them as well as on
$L$ as far as $L \agt \lambda/2$.  For model C, the chosen parameters
are (200, 120, 30, 6, 1/12, 0.53), (220, 125, 30, 6, 9/120, 0.58),
(220, 125, 30, 7, 8/120, 0.58), and (220, 140, 30, 8, 7/120, 0.48).

\begin{figure}[t]
\begin{center}
\epsfxsize=3.in
\leavevmode
\epsffile{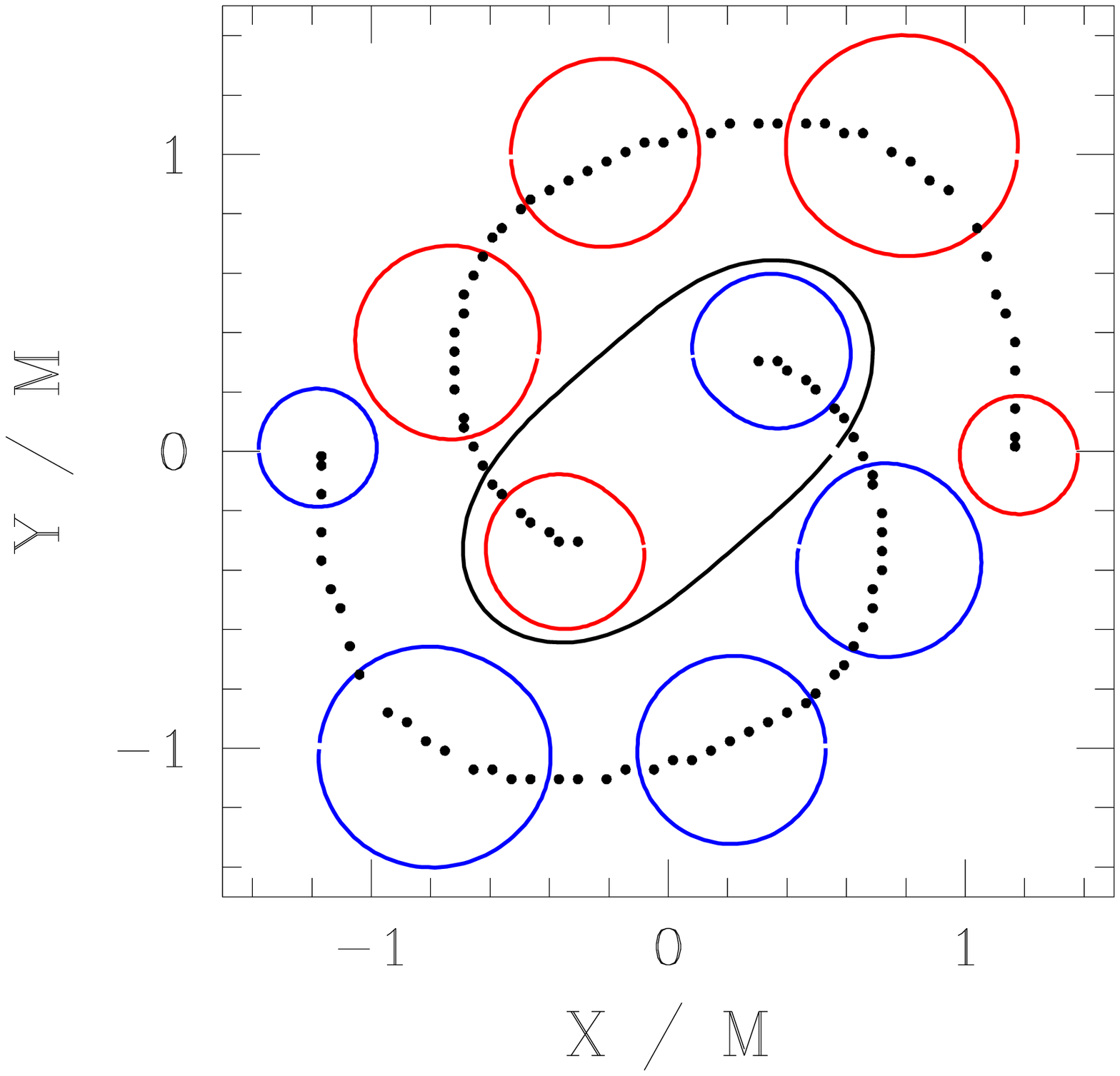}
\epsfxsize=3.in
\leavevmode
\epsffile{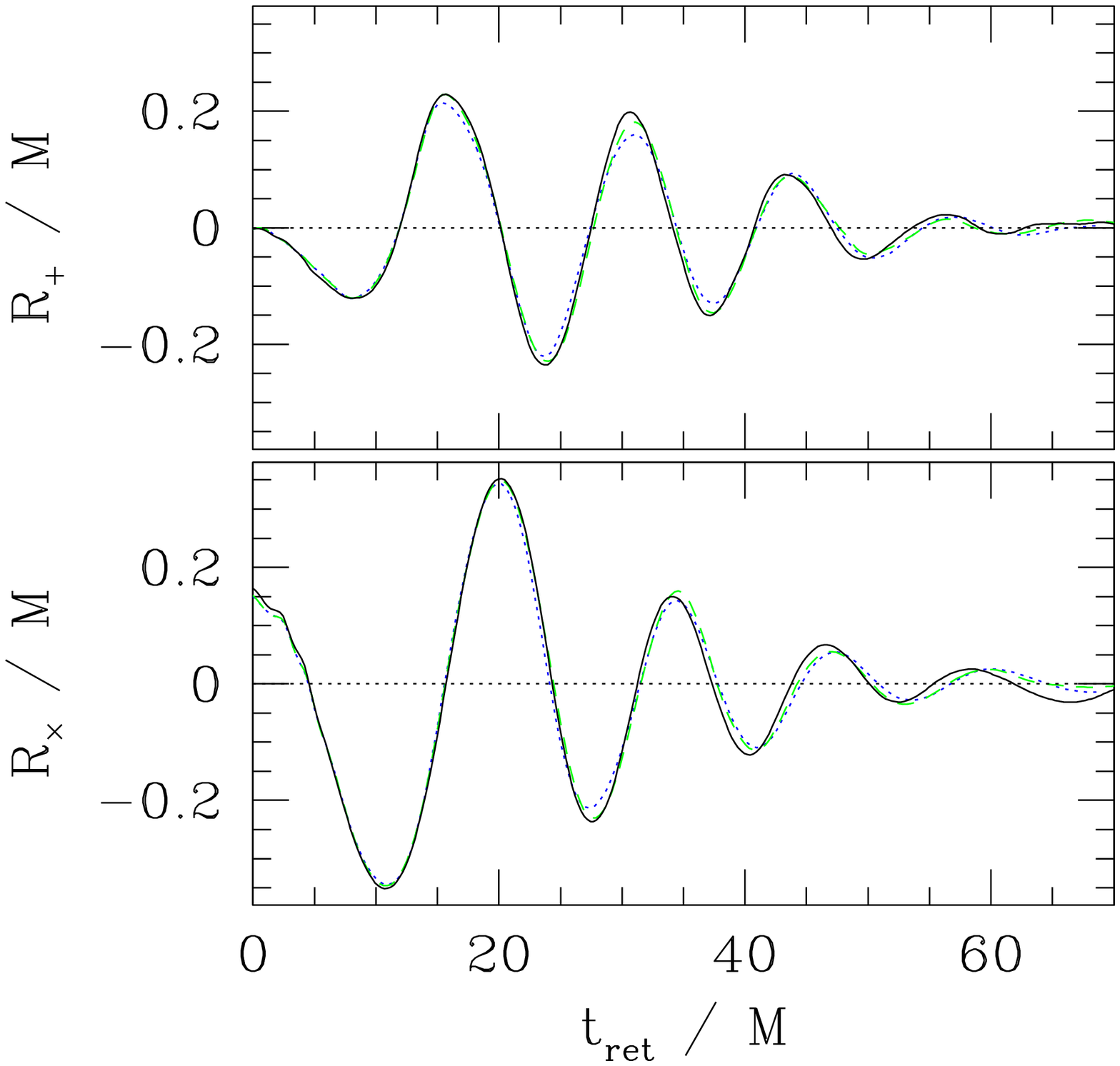}
\end{center}
\vspace{-10mm}
\caption{Numerical results for merger of binary BHs performed with
the initial data of \cite{BB2}. Left: The open thick circles denote
evolution of the location of the apparent horizon of two BHs for
$t/M=0$, 5, 10, 15, and 20 where $M$ is the total ADM mass at $t=0$.
The open wide circle at $t=20M$ is the 
common apparent horizon. The small solid circles denote approximate
location of the maximum of the conformal factor. The grid setting is
$(N, N_0, \Delta i, \xi, \Delta x/M_p, L/\lambda)=$
(200,105,30,7,1/16,1.20) for this result (note that $\lambda \approx 37M$). 
Right: $+$ and $\times$ modes of
gravitational waveforms extracted $r \approx \lambda$
for $(N, N_0, \Delta i, \xi, \Delta x/M_p, L/\lambda)=$
(200,105,30,9,0.05,1.20) (solid curves), (200,105,30,7,0.064,1.23)
(dashed curves), and (200,105,30,5,0.08,1.23) (dotted curves).
\label{FIG0}}
\end{figure}

For a test, we performed simulations for merger of two nonspinning BHs
adopting the same initial condition as used in \cite{BB2}. We
focused particularly on the merger time, which is referred to as the
time at which a common apparent horizon is first formed and found that
it varies with improving grid resolution approximately at 1st
order. The likely reason is that geometric variables vary steeply
around the BH where they are evolved with 1st-order accuracy in our
scheme, although other regions are resolved with 2nd-order accuracy. 
By the extrapolation, an exact merger time is estimated to be $\approx 19M$.
This result agrees approximately with those of \cite{BB2}.  This
indicates that our code can follow moving punctures 
(see the left panel of Fig. \ref{FIG0} for evolution of the location
of apparent horizons for the initial condition of \cite{BB2}).

Gravitational waves are also computed. In the right panel of
Fig. \ref{FIG0}, we display the results for three different grid
resolutions. The figure shows that after merger, gravitational waves
are determined by the quasinormal mode ringing and that the waveforms
depend very weakly for chosen grid resolutions. The wavelength of the
quasinormal mode is $\approx 11$--12M, which agrees with the result of
\cite{BB2}. 

\begin{figure*}[t]
\begin{center}
\epsfxsize=2.5in
\leavevmode
\epsffile{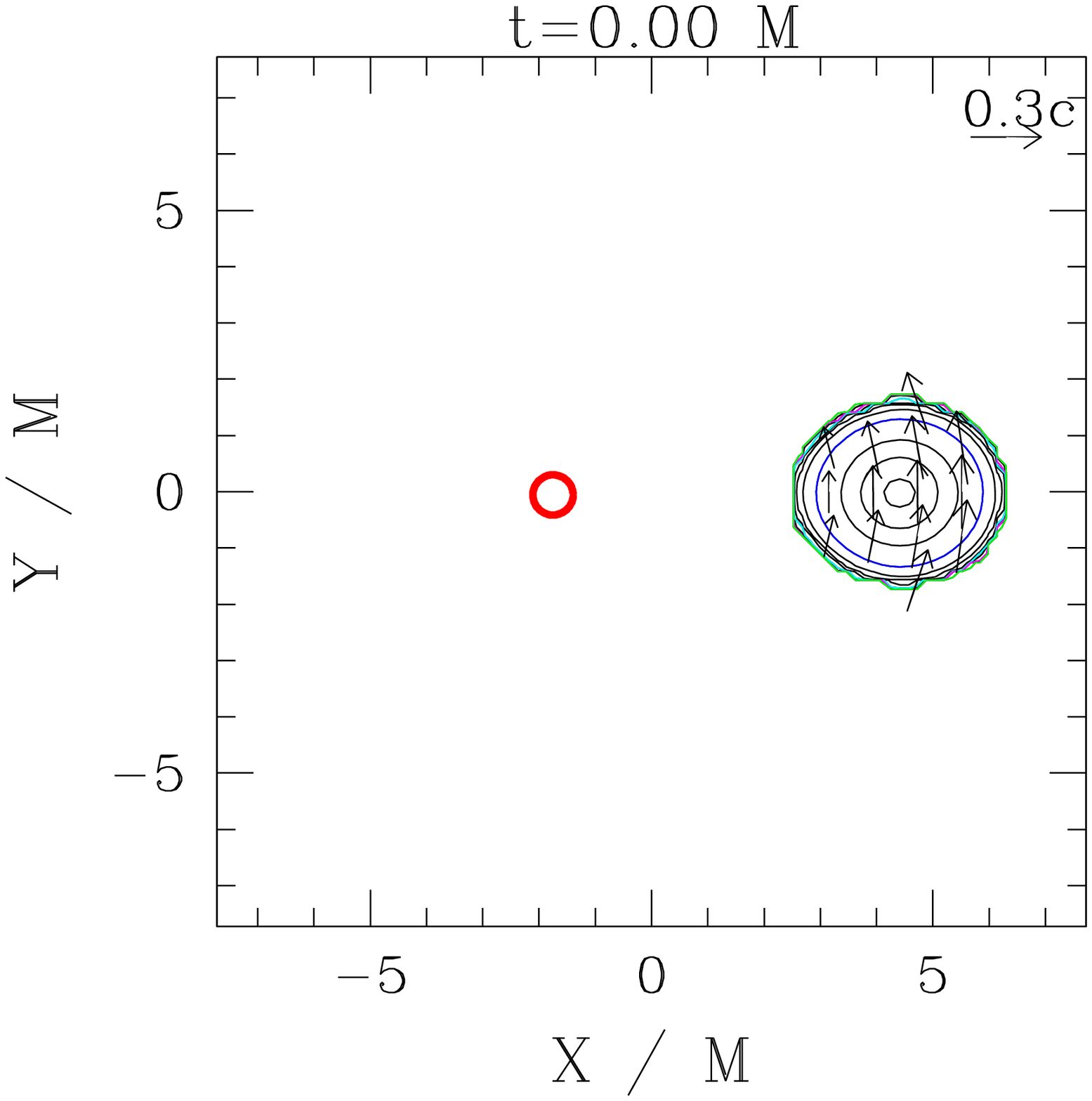}
\epsfxsize=2.5in
\leavevmode
\hspace{-1.95cm}\epsffile{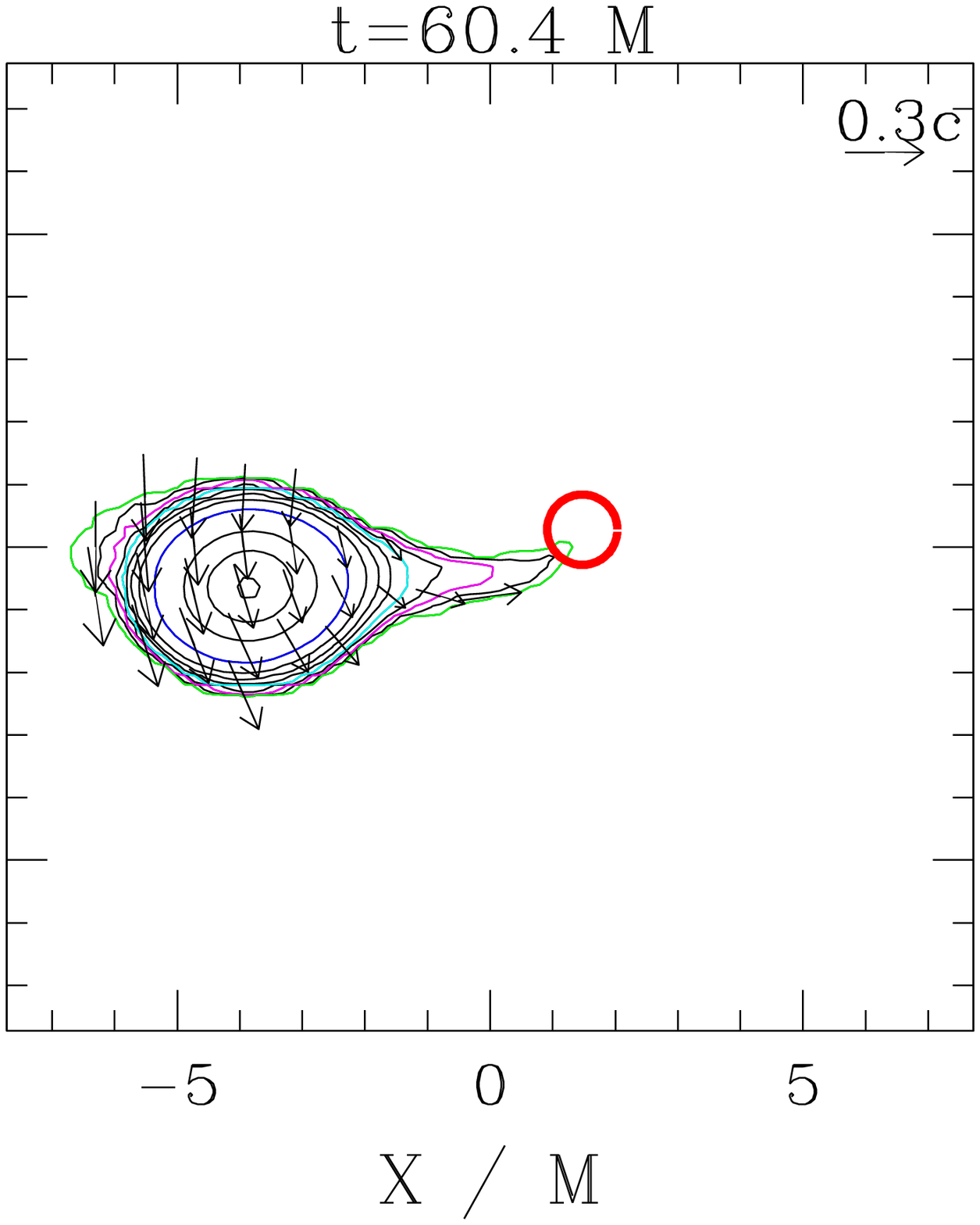} 
\epsfxsize=2.5in
\leavevmode
\hspace{-1.95cm}\epsffile{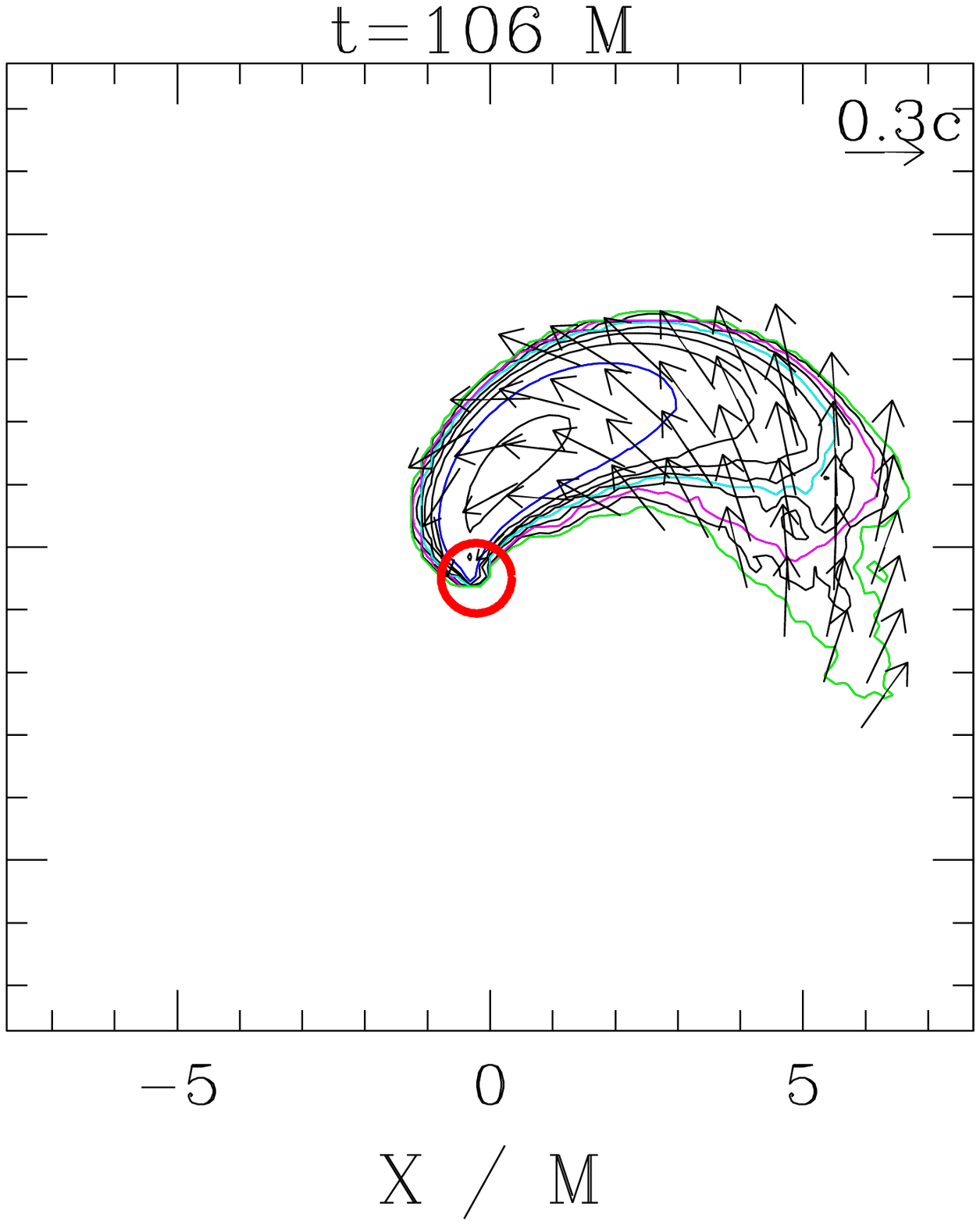} \\
\vspace{-7mm}
\epsfxsize=2.5in
\leavevmode
\epsffile{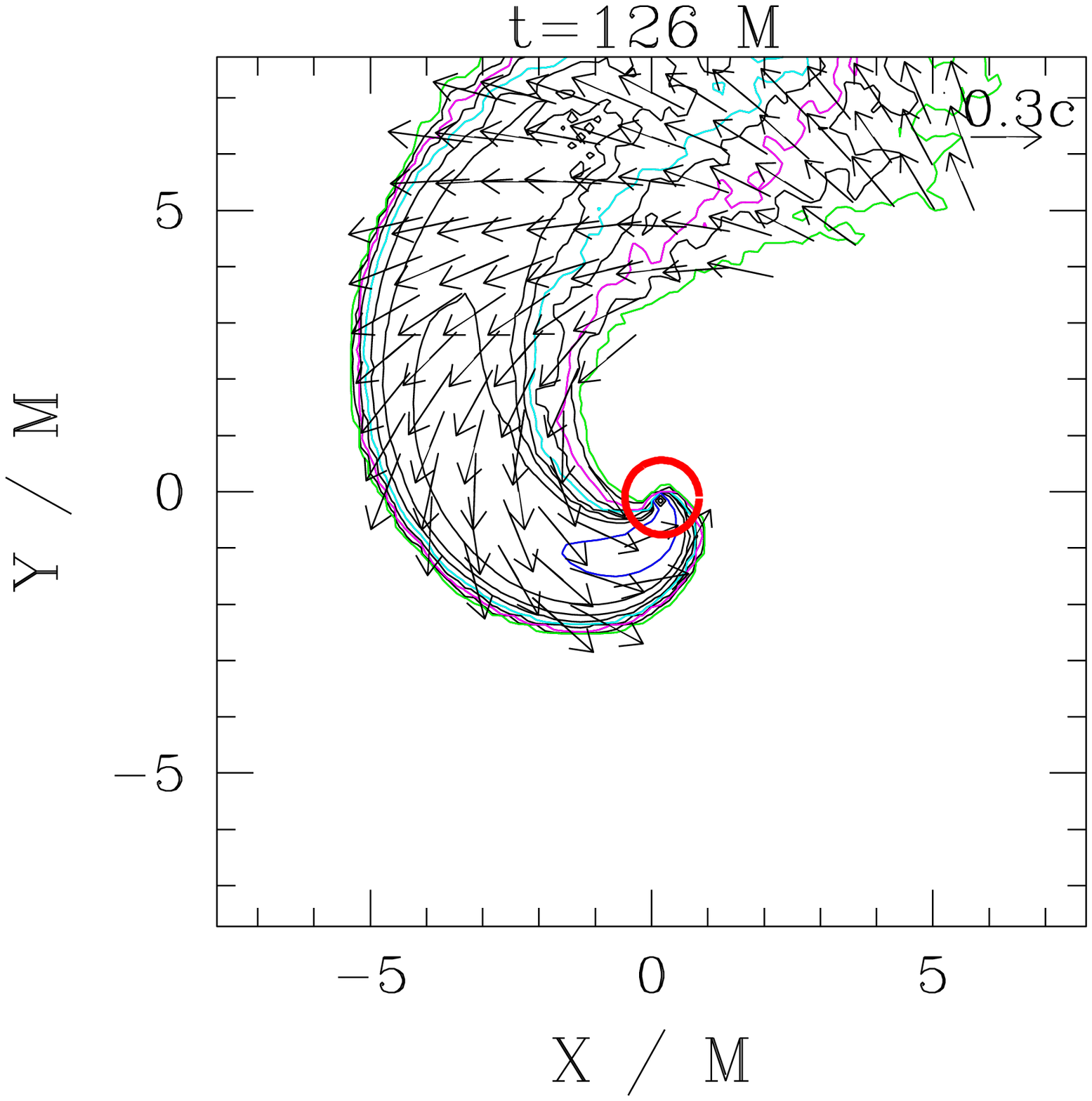}
\epsfxsize=2.5in
\leavevmode
\hspace{-1.95cm}\epsffile{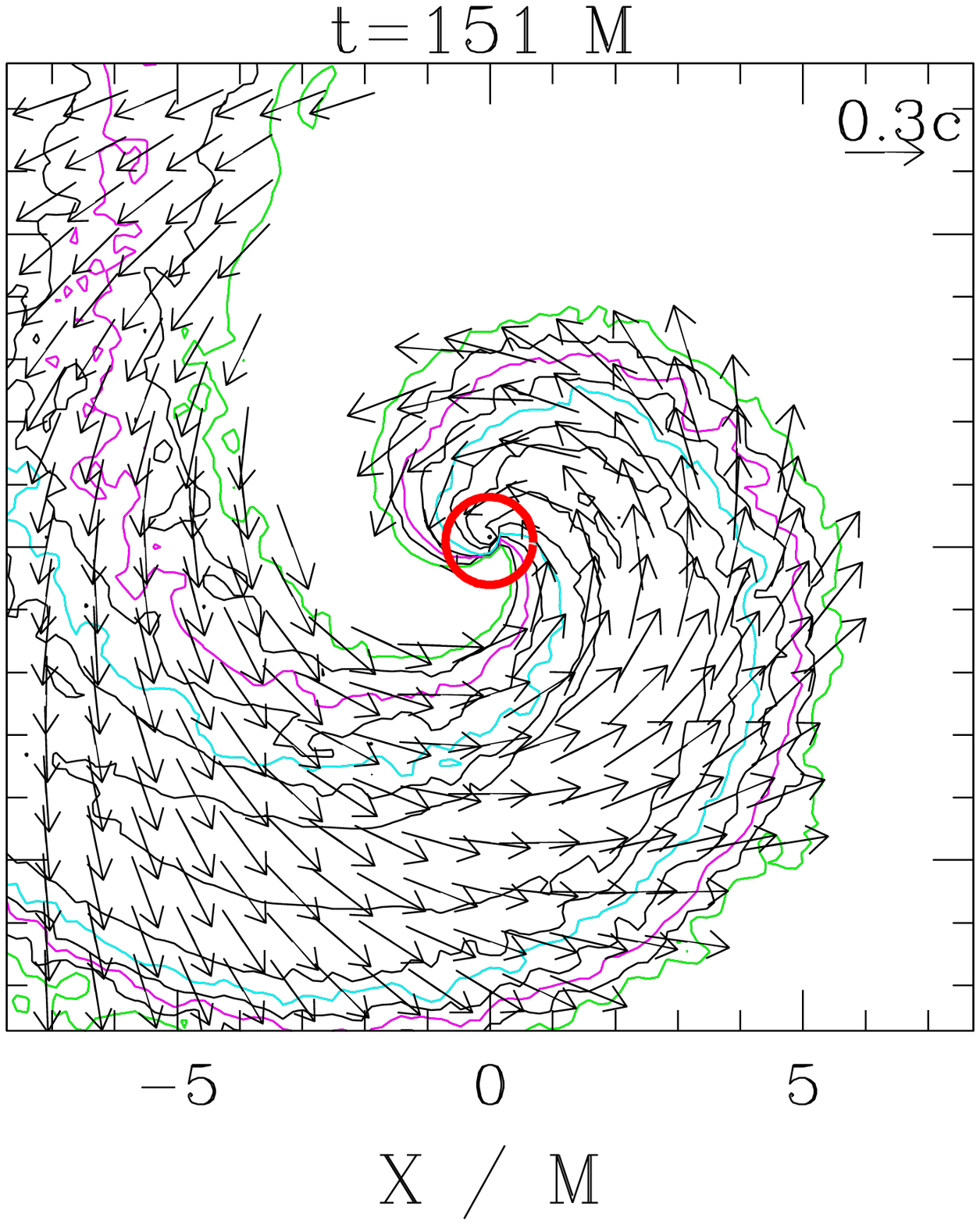}
\epsfxsize=2.5in
\leavevmode
\hspace{-1.95cm}\epsffile{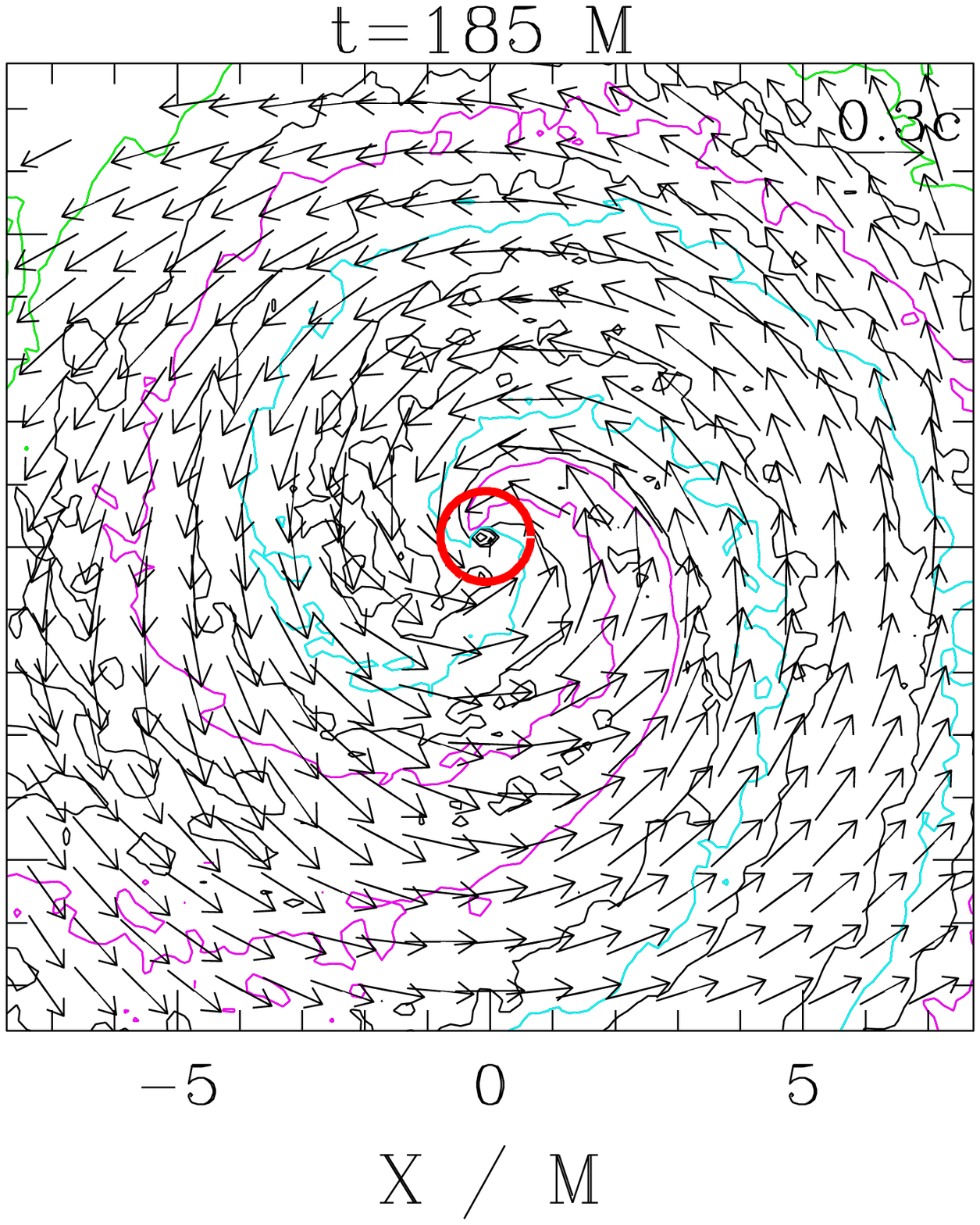}
\vspace{-6mm}
\caption{Snapshots of the density contour curves for $\rho$ in the
equatorial plane for model A with
$(N, N_0, \Delta i, \xi, \Delta x/M_p, L/\lambda)=$(220,150,9,3/40, 0.59). 
The solid contour curves are drawn for
$\rho= (1+2i)\times 10^{14}~{\rm g/cm^3}~(i=1, 2, 3)$ and
for $10^{14-0.5 i}~{\rm g/cm^3}~(i=1 \sim 8)$. 
The maximum density at $t=0$ is $\approx 7.2 \times 10^{14}~{\rm g/cm^3}$. 
The blue, cyan, magenta, and green curves denote
$10^{14}$, $10^{12}$, $10^{11}$, and $10^{10}~{\rm g/cm^3}$, respectively.
Vectors indicate the velocity field $(v^x,v^y)$, and the scale is shown in the
upper right-hand corner. The thick (red) circles are apparent horizons.
Time is shown in units of total mass of the system $M$.
\label{FIG1}}
\end{center}
\end{figure*}

Next, we present the results for merger of BH and NS for model A. 
Figure \ref{FIG1} shows evolution of contour curves for $\rho$ and
velocity vectors for $v^i$ in the equatorial plane together with the
location of apparent horizons at selected time slices for $(N, N_0,
\Delta i, \xi, \Delta x/M_p, L/\lambda)=$(220,150,9,3/40, 0.59). Due 
to gravitational radiation reaction, the orbital radius decreases and
then the NS is elongated (2nd panel). Because of the elongation, the
quadrupole moment of the NS is amplified and the attractive force
between two objects is strengthen \cite{LRS}. This effect accelerates
an inward motion and, consequently, the NS starts plunging to the BH
at $t \sim 90M$. Soon after this time, the NS is tidally disrupted;
but the tidal disruption occurs near the ISCO and hence the material
in the inner part is quickly swallowed into the BH (3rd and 4th
panels). On the other hand, because of the outward angular momentum
transfer, the material in the outer part of the NS forms a disk with
the maximum density $\sim 10^{12}~{\rm g/cm^3}$ (5th and 6th panels). 

\begin{figure}[t]
\begin{center}
\epsfxsize=3.in
\leavevmode
\epsffile{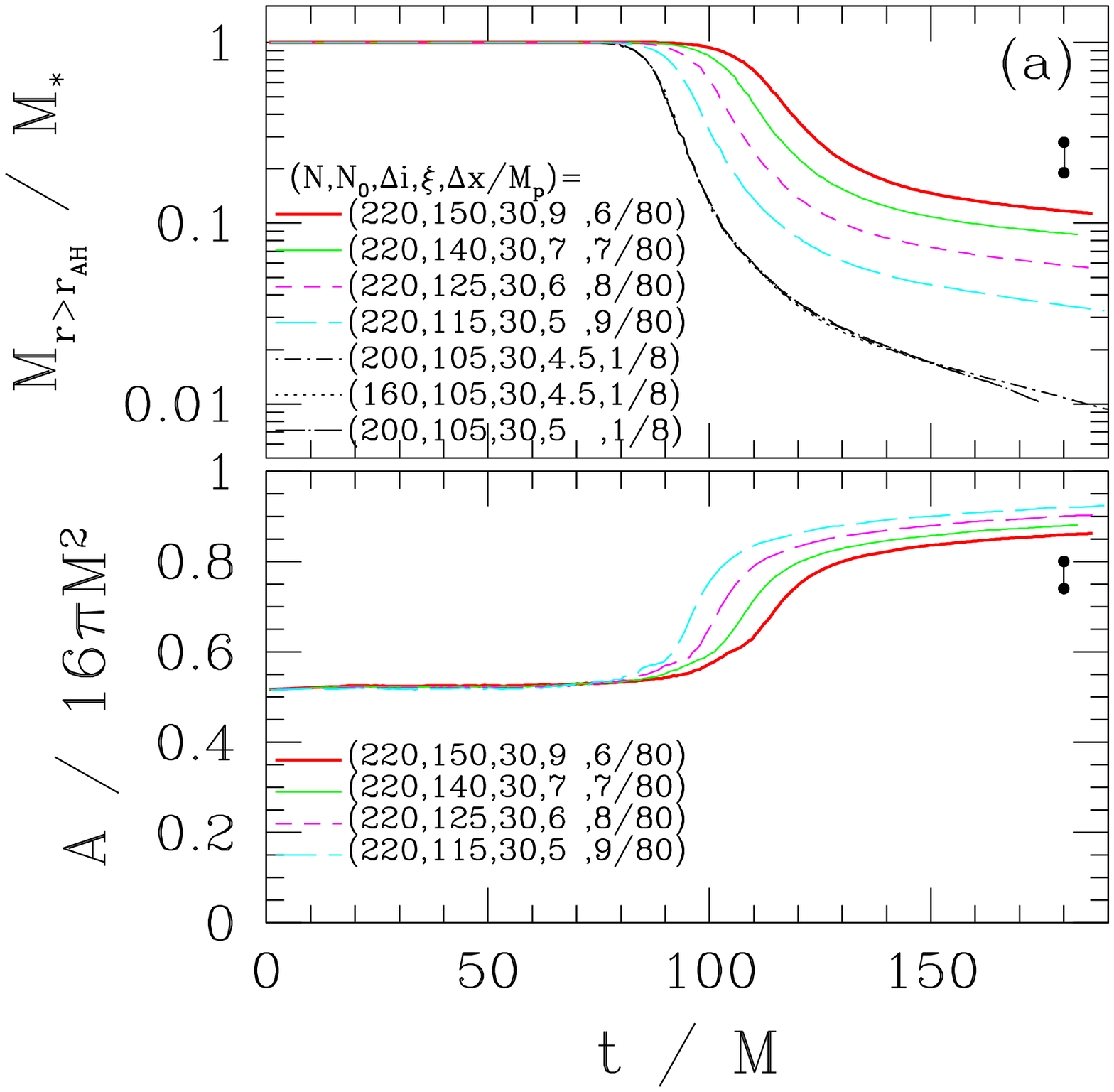}
\epsfxsize=3.in
\leavevmode
\epsffile{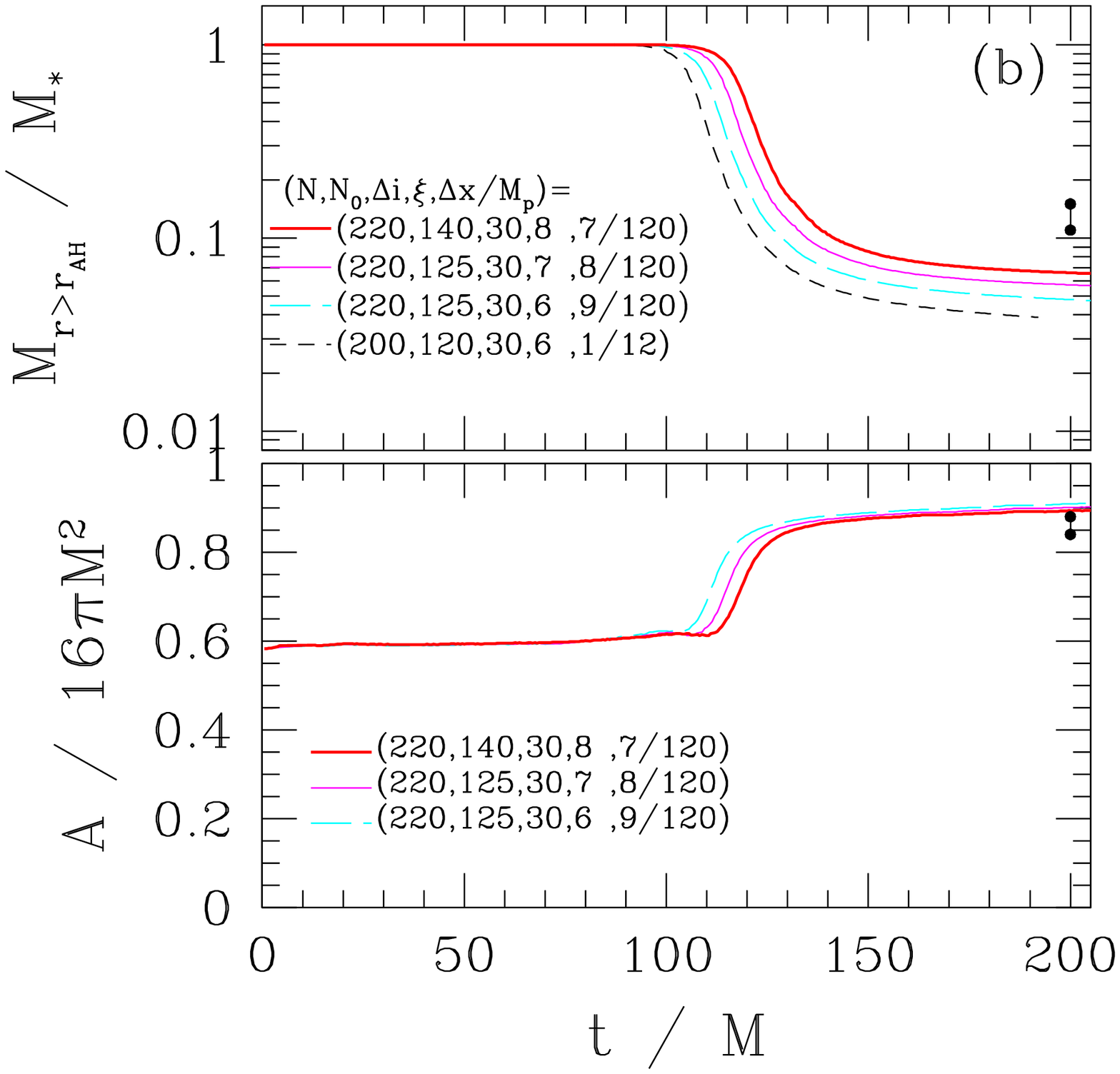}
\end{center}
\vspace{-6mm}
\caption{(a) Results for model A for various grid resolutions.  Upper
panel shows evolution of baryon rest-mass located outside the apparent
horizon. The plots for $\Delta x=M_{\rm P}/8$ almost coincide and show
that the results depend weakly on the values of $L$ and $\xi$. On the
other hand, the results depend systematically on $\Delta x$ (see
text). Lower panel shows evolution of area of apparent horizon in
units of $16\pi M^2$. (b) The same as (a) but for model C.
\label{FIG2}}
\end{figure}

Mass of the disk is not large. Figure \ref{FIG2}(a) shows evolution of
baryon rest-mass located outside apparent horizons $M_{r>r_{\rm AH}}$.
We find that $\sim 80\%$ and $\sim 90\%$ of the mass is swallowed into
the BH in $t \sim 130M \sim 2$ ms for models A and C, respectively,
for the best grid-resolution simulations.  The swallowing continues
after this time. Note that since $M_{r > r_{\rm AH}}$ is rest-mass
outside apparent horizons, the disk mass which should be defined for
the mass located outside an ISCO around the formed BH is slightly
smaller. We follow the evolution of the rest-mass of material located
for $r >3M$ and $r > 4.5M$, where $r=3M$--$4.5M$ are approximate
locations of the ISCO around the formed BH, and find that their values
are smaller than $M_{r > r_{\rm AH}}$ by $\sim 10\%$ and 20\%,
respectively. Thus, the disk mass would be 0.8--$0.9M_{r > r_{\rm
AH}}$ in reality.

The value of $M_{r > r_{\rm AH}}$ depends systematically on $\Delta
x$; we find that the results for good resolutions ($\Delta
x/M_p=3/40$, 7/80, and 1/10 for model A and 7/120, 8/120, and 9/120
for model C) at late times approximately obey a relation of
convergence, i.e., $M_{r > r_{\rm AH}}(t)=a(t)+b(t)\Delta x^n$ where
$a(t)$ and $b(t)$ are functions of time. The order of convergence,
denoted by $n$, is between 1st and 2nd order (i.e., $1 < n < 2$). For
model A, least-square fitting gives $a(t)$ at $t=180M$ as $\approx
0.19M_*$ if we set $n=2$ and as $0.28M_*$ for $n=1$ (see the solid
circle in Fig. \ref{FIG2}(a)). For model C, they are $0.11M_*$ and
$0.15M_*$ (the solid circle in Fig. \ref{FIG2}(b)),
respectively. Thus, the true result should be between $0.19M_*$ and
$0.28M_*$ for model A and between $0.11M_*$ and $0.15M_*$ for model C.

It is reasonable that the disk mass for model C is smaller than for 
model A since the mass ratio of NS to BH is smaller for model C while 
the compactness of NS is nearly identical. A remarkable point is that with a
small increase of the BH mass from $3.2M_{\odot}$ to
$4.0M_{\odot}$, the disk mass decreases by a factor of 2. This
suggests that the disk mass depends sensitively on the BH mass. 

Note that the adopted NS has corotating velocity field initially and
furthermore its radius is larger than that of canonical NSs. For
irrotational velocity field with a realistic value of radius, the disk
mass would be smaller than this value. It is reasonable to consider
that our results for the disk mass provide an upper limit of the disk
mass for given mass and spin of BH.  Hence it is unlikely that a
massive disk with $\sim M_{\odot}$ is formed after merger of
nonspinning BH of mass $M > 3M_{\odot}$ and canonical NS of mass
$\approx 1.4M_{\odot}$ and radius $\approx 11$--13 km, although a disk
of mass of $\sim 0.2$--$0.3M_{\odot}$ may be formed for small BH mass. 
This value of the disk mass is large enough to explain short GRBs of
relatively low total energy $\sim 10^{49}$ ergs (e.g., \cite{RJ} for
an estimate). 

Lower panels of Fig. \ref{FIG2} show the evolution of the area of
apparent horizons in units of $16\pi M^2$. These illustrate
that the area of BHs quickly increases swallowing material, and then, 
the area settles down to an approximate constant. We, again, evaluate
the true final area by extrapolation for results with 
different grid resolutions.  For model A, the area in units of $16\pi
M^2$ at $t=180M$ is 0.74 and 0.80 in the assumption of the 1st ($n=1$)
and 2nd-order ($n=2$) convergences, respectively (see the solid
circles in Fig. \ref{FIG2}(a)). For model C, $A/16\pi M^2$ at $t=200M$
is 0.84 and 0.88 for $n=1$ and $n=2$, respectively (see the solid
circles in Fig. \ref{FIG2}(b)). From these values the spin parameter
of formed BHs, $a$, is approximately derived from
\beqn
{A \over 16\pi M_{\rm BHf}^2}={1 + \sqrt{1-a^2} \over 2},
\eeqn
where $M_{\rm BHf}$ denotes the mass of the formed BH.  To 
approximately estimate it, we simply use
\beqn
M_{\rm BHf}=M-M_{r> r_{\rm AH}}-E_{\rm GW},
\eeqn
where $E_{\rm GW}$ is radiated energy by gravitational waves. We find that
$E_{\rm GW}$ is about 1\% of $M$ and simply use relation $E_{\rm
GW}=0.01M$.  Then we obtain $a=0.57$ and 0.52 for $n=1$ and 2 for
model A and $a=0.52$ and 0.42 for $n=1$ and $n=2$ for model C. Thus, 
spinning BHs of moderate rotation are outcomes. 

The spin parameter of the formed BHs is much smaller than the initial
value of $J/M^2$ of the system. One of the reasons is that the disk
has large angular momentum approximately written as $3M_{\rm
BHf}M_{\rm disk}$ where $M_{\rm disk}$ denotes the disk mass $\sim
0.8$--$0.9M_{r > r_{\rm AH}}$ Here, the factor $3M_{\rm BHf}$ denotes
a value of typical specific angular momentum around the formed
BH. Denoting the initial angular momentum by $a_0 M^2$ where $a_0
\approx 0.73$ and 0.65 for models A and C (see Table I), the fraction
of the angular momentum that the disk has is $\sim 3a_0^{-1} M_{\rm
disk}M_{\rm BHf}/M^2$. Thus, for model A, the fraction is $\sim
20$--30\% and for model C, it is 10--15\%. In addition, gravitational
waves carry away the angular momentum by $\sim 10\%$ of
$a_0M^2$. Thus, the angular momentum of the formed BH should be
smaller than the initial total angular momentum of the system by
30--40\% for model A and by 20--25\% for model C. Therefore, the
values for $a$ derived above are reasonable magnitudes. 

\begin{figure}[t]
\begin{center}
\epsfxsize=3.3in
\leavevmode
\epsffile{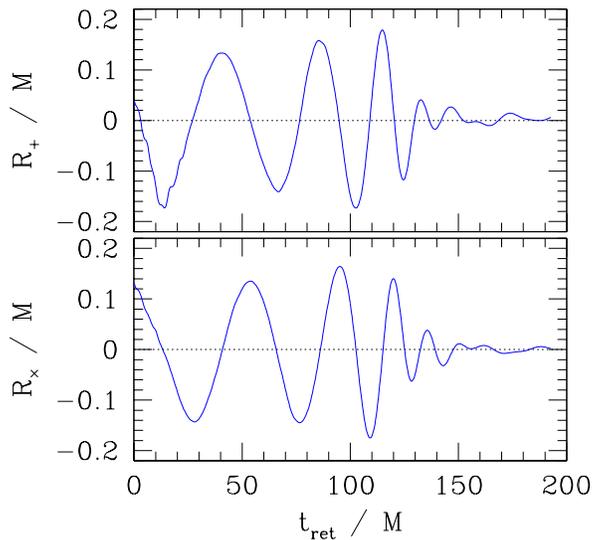}
\end{center}
\vspace{-9mm}
\caption{$+$ and $\times$ modes of gravitational waveforms for model C.
$t_{\rm ret}$ and $M$ denotes the retarded time and the ADM mass of
the system, respectively. The amplitude at a distance of observer can
be found from Eq. (\ref{hamp}). 
\label{FIG3}}
\end{figure}

In Fig. \ref{FIG3}, gravitational waveforms for model C is shown.
Gravitational waves are extracted from the metric near the
outer boundaries using a gauge-invariant wave extraction method (see
\cite{STU0,SN} for details). From the values of $R_+$ and $R_{\times}$, the 
maximum amplitude of gravitational waves at a distance $D$ is evaluated
\beqn
h_{\rm gw} \approx 10^{-22} \biggl( {\sqrt{R_{+}^2+R_{\times}^2}
\over 0.31~{\rm km}}\biggr)
\biggl({100~{\rm Mpc} \over D}\biggr). \label{hamp}
\eeqn
Here, the maximum amplitude can be observed if the observer is located along 
the $z$-axis. Figure \ref{FIG3} implies that the maximum amplitude at
a distance of $D=100$ Mpc is $\approx 5 \times 10^{-22}$ since 
$M=5.26M_{\odot}$. 

For $t_{\rm ret} \alt 120M$, inspiral waveforms are seen: Amplitude
increases and characteristic wavelength decreases with time. The
wavelength at the final phase of the inspiral is $\sim 25M$ indicating
that the orbital period of the ISCO is $\sim 50M$ (i.e., the angular
velocity is $\sim 0.12M^{-1}$).  This is in approximate agreement with
the 3rd post-Newtonian results \cite{Luc}.

For $120M \alt t_{\rm ret} \alt 150M$, ring-down waveforms are
seen. The characteristic wavelength is $\sim 15M_{\rm BHf}$,
which is in approximate agreement with the wavelength of
the quasinormal mode. 

The characteristic feature of gravitational waves after the merger 
sets in is that the amplitude damps quickly even in the
formation of massive disk. This is due to the fact that the degree of 
nonaxisymmetry of the disk decreases  in a very short time
scale ($\sim 20$--$30M$). This indicates that in the frequency
domain, the amplitude of Fourier power spectrum steeply decreases in
the high-frequency region.

\begin{figure}[t]
\begin{center}
\epsfxsize=3.3in
\leavevmode
\epsffile{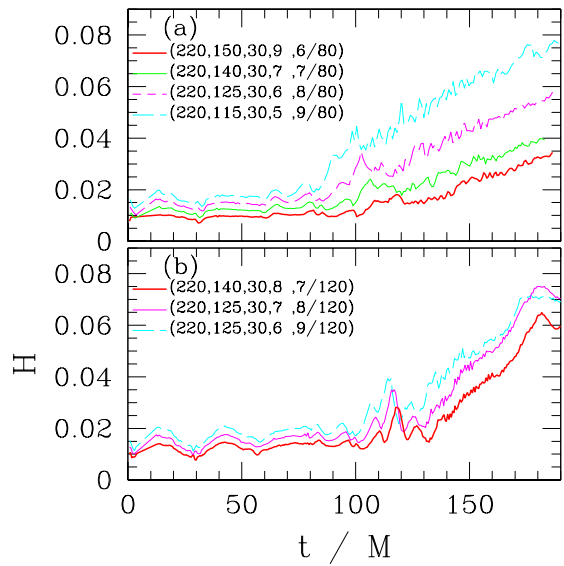}
\end{center}
\vspace{-9mm}
\caption{Evolution of averaged violation of the Hamiltonian constraint
(a) for model A and (b) for C. 
\label{FIG4}}
\end{figure}

Figure \ref{FIG4} shows the evolution of averaged violation of the
Hamiltonian constraint. For the average, rest-mass density is used as
a weight (see \cite{STU0} for definition) and the integral is
performed for the region outside apparent horizons. The figure shows
that the Hamiltonian constraint converges approximately at 2nd
order. This result is consistent with the fact that the region except
for the vicinity of BH is followed with 2nd-order accuracy.

\section{Discussion}

In this paper, we have presented our latest numerical results of fully 
GR simulation for merger of BH-NS binary, focusing on the case that
the BH is not spinning initially and the mass ratio $q$ is fairly
large as 0.3--0.4. It is found that even with such high values of $q$, the
NS is tidally disrupted only for the orbit very close to ISCO and 
80--90\% of the mass element is quickly swallowed into the BH
without forming massive disks.  The results do not agree
quantitatively with the prediction by the tidal approximation
study. The reasons are: (1) In the tidal approximation, one describes
NSs by the Newtonian gravity. In general relativity, gravity is
stronger and the tidal disruption is less likely. (2) The time scale
for angular momentum transfer during tidal disruption near the ISCO is
nearly as long as the plunging time scale determined by gravitational
radiation reaction and attractive force between two objects. Hence
before the tidal disruption completes, most of the material is
swallowed. 

If the BH has a large spin, the final fate may be largely changed
because of the presence of spin-orbit repulsive force. This force can
weaken the attractive force between BH and NS and slow down the
orbital velocity, resulting in smaller gravitational wave luminosity
and longer radiation reaction time scale \cite{KWW}.  This effect may
help massive disk formation. The study of spinning BH binaries is one
of the next issues. The fate will also depend on EOS of NS \cite{ISM}
and mass of BH as well as on the velocity field of NS. Simulations with
various EOSs and BH mass and with irrotational velocity field
are also the next issues.

In this paper, a small number of the results for quasicircular states
of BH-NS binaries are presented. Currently we are working in the
computation of quasicircular states for a wide variety of masses
of BH and NS in the framework described in Sec. 2. The numerical
results will be reported in the near future \cite{US07}. 


\noindent
\underline{{\em Acknowledgements}}: 
MS thanks Y. Sekiguchi for useful discussion. 
Numerical computations were performed on the FACOM-VPP5000 at ADAC at
NAOJ and on the NEC-SX8 at YITP in Kyoto University. This work was 
supported in part by Monbukagakusho Grants (No. 17540232)
and by NSF grants PHY0071044, 0503366.

\end{document}